\documentclass[%
onecolumn,%
oneside,%
floats,%
aps,%
 prd,%
nobibnotes,%
nofootinbib,%
amsmath,%
amssymb,%
amsfonts,%
amscd,%
  superscriptaddress,%
eqsecnum%
]{revtex4} 

\usepackage[utf8]{inputenc}
\usepackage{graphicx,array,dcolumn}
\usepackage{cases}
\usepackage[paperwidth=210mm,paperheight=297mm,centering,hmargin=1.8cm,vmargin=2.5cm]{geometry}
\usepackage{enumerate}

\usepackage{hyperref}
\hypersetup{hidelinks}

\newcommand{\bi}{\bibitem}

\def\be{\begin{eqnarray}}
\def\ee{\end{eqnarray}}

\def\-g{\sqrt{-g}}

\renewcommand\rho{\varrho}

\begin{document}

\title{
 Antimatter in the universe and Milky Way
}

\author{A.D. Dolgov}
\email{dolgov@nsu.ru}
\affiliation{Novosibirsk State University, Novosibirsk, 630090, Russia}
\affiliation{Bogolyubov Laboratory of Theoretical Physics, JINR, Dubna, 141980, Russia}

\begin{abstract}

Theoretical model leading to a significant Galaxy population by primordial antimatter is described.
Observational data confirming this "crazy" prediction are presented.

\end{abstract}

\maketitle

 \section{Introduction}\label{s-intro}

The conventional cosmological faith rquired that each galaxy consists only of matter and any 
amount of antimatter should be of secondary origin, created via cosmic ray interactions, and 
hence negligibly tiny. 

The first person who doubted this seemingly evident "fact" was  B.P. Konstantinov with 
coauthors~\cite{konst1,konst2,konst3}. They searched for meteors consisting of antimatter, 
despite being very strongly criticized by Ya.B. Zeldovich, who was in very good personal relations 
with Konstantinov, but Zeldovich was probably acting  according to his credo: "Amicus Plato, sed magis 
amica veritas" - "Platon is my friend but the truth  is dearer to me".

However, a very strong evidence is now obtained, confirming existence  of
different type and noticeable population of Milky Way with antimatter thus supporting
Konstantinov's "heresy".   

The notion of antimatter appeared at the end of  the XIX century in the publication of the 
British physicist Arthur Schuster.  In 1898, one year after discovery of electron (J.J. Thomson, 1897) 
he published a paper~\cite{schuster} with the
conjecture that there might be other sign electricity, antimatter, and supposed that there 
might be entire solar systems, made of antimatter, indistinguishable from ours. Schuster 
made fantastic wild guess that matter and antimatter are capable to annihilate and produce 
vast energy. However, he believed that they were gravitationally repulsive having negative mass. 
Schuster concluded that two such objects on close contact should have vanishing mass!
Quoting his paper~\cite{schuster}: “When the year’s work is over and all sense of responsibility 
has left us, who has not occasionally set his fancy free to dream about the unknown, perhaps 
the unknowable?’... Astronomy, the oldest
and yet most juvenile of the sciences, may still have some surprises in store. May antimatter be 
commended to its case”.

Except for an incorrect assumption of negative mass of antimatter particles, the rest of Schuster's 
work  was ingenuous guess ahead of scientific development for almost a half of century. 
Still the notion of negative mass  is revived even nowadays by some illiterate authors.

The real scientific father of antimatter is definitely Paul Dirac. 
According to Werner Heisenberg: “his
discovery of antimatter was perhaps the biggest jump
of all big jumps in physics in our century”~\cite{heisenberg}. 

In his Nobel Lecture at December 12, 1933 “Theory of electrons and positrons” 
dedicated to his prediction 
of positrons, Dirac anticipated that there could be antistars and
possibly antiworlds: ”... It is quite possible that... these stars being built up mainly of 
positrons and negative
protons. In fact, there may be half the stars of each kind. The two kinds of stars would both show 
exactly the same spectra, and there would be no way of distinguishing them by present 
astronomical methods.”

Strictly speaking the last statement is not correct. There are some, though difficult, may be unrealistic,  
ways to distinguish a star from an antistar even at very large distances, as is found in ref.~\cite{DNV}.
the spectra of the emitted radiation are not exactly the same and the polarization of radiation from 
weak decays
could be a good indicator and lastly the spectrum of emitted neutrinos versus antineutrinos from 
supernovae or anti-supernovae would be different. In the exotic case of CPT violation, when masses 
of particles and antiparticles
and the decay probabilities are not the same, more interesting possibilities would be open.

\section{Galaxies and anti-galaxies}

\subsection{Pioneering papers}

Possible existence of anti-galaxies, i.e. galaxies consisting solely of antimatter was foreseen by
Floyd Stecker with collaborators in 1971~\cite{stecker-1} and later was further developed in his works 
of 1981~\cite{stecker-2,stecker-3}.
The authors of ref.~\cite{stecker-1} argued that according to 
the results of their calculation of the cosmological $\gamma$-ray spectrum from matter-antimatter 
annihilation in the universe, the similarity of the calculated spectrum with the present observations 
of the $\gamma$-ray background spectrum above 1 MeV suggests that such observations may be 
evidence of the existence of antimatter on a large scale in the universe.

It is shown in refs.~\cite{stecker-3,stecker-2} that in the 
framework of grand unified theories, spontaneous CP violation leads to a domain 
structure in the universe 
with the domains evolving into separate regions of matter and antimatter excesses. 
Subsequent to exponential horizon growth, this can result in a universe of matter galaxies and 
antimatter galaxies. 
Various astrophysical data appear to favor this form of big bang cosmology. 
Essentially the same ideas were replicated a year later in ref.~\cite{khlop-1}.

A review  of "anti-situation" in the state of art at  2002 is presented 
in two keynote lectures at 14th  
Rencontres de Blois on Matter - Anti-matter Asymmetry~\cite{steck-rev,ad-rev}.


 \subsection{Matter and antimatter in the Universe \label{anti-universe}}
 
  To the best of my knowledge the first papers on cosmological antimatter were published 
 by F. Stecker in 1971~\cite{stecker-1}, independently on the 
 three year earlier papers by Konstantinov et al~\cite{konst1,konst2,konst3} 
 on search of antimatter in the Galaxy.
  Further development of the idea of matter-antimatter 
 domain structure of the universe was presented in ref.~\cite{stecker-2}.
 
The analysis, performed  in ref.~\cite{C-AdR-G}, permits to conclude that matter–antimatter 
symmetric universe 
or close to the symmetric state, is excluded by the observed
 cosmic diffuse gamma-ray background and a distortion of the cosmic microwave background,
 if the antimatter domains are not sufficiently far away.
 
However, there still remains some space for the fraction of cosmological antimatter, but
considerably restricted. It is argued by G. Steigman~\cite{GS-1} in 1976,
that the nearest anti-galaxy should be out of our galaxy cluster and thus could not
be closer than at $\sim10$ Mpc.
In a subsequent paper by the same author~\cite{GS-2} it is shown that
the fraction of antimatter in Bullet Cluster should be below  $ 3\times 10^{-6}$. 

 \section{Bounds on antimatter in Milky Way \label{ss-anti-galaxy}}
 
As mentioned above, the search of antimatter in the Milky Way was intiiated 
 by Konstantinov with coworkers in 1968~\cite{konst1,konst2,konst3}.
 Until recently there was no reason to suspect that any 
 noticeable amount of antimatter might be in the Galaxy. The predictions of galactic antimatter made
the papers~\cite{DS,DKK} were not taken seriously.
However, now there are plenty  of data indicating that Milky Way contains significant amount 
of antimatter of different kinds: positrons, antinuclei, 
and possibly antistars.  The observations do not violate the existing  bounds on galactic antimatter. 
According to the results of refs.~\cite{DS,DKK} antimatter objects could be not only in the 
Galaxy but in its halo as well.

According to ref.~\cite{ballmoos-bound}, the analysis of the intensity of gamma rays 
created by the Bondi
accretion of interstellar gas to the surface of an antistar 
allows to put a limit on the relative density of antistars in the Solar neighbourhood: $N_{\bar *} /N_* < 4 \cdot 10^{-5} $
inside 150 pc from the Sun.

The bounds on galactic antimatter are analysed in refs.~\cite{BD,db,bdp}. 
The limits on the density of galactic antistars are rather loose, because the annihilation proceeds 
only on the surface of antistars, i.e on the objects with short mean free path of protons, so the extra 
luminosity created by matter-antimatter annihilation is relatively low. 

\section{Galactic antimatter prediction \label{s-anti-prediction}}

\subsection{Supporting data \label{ss-support}}

In papers\cite{DS,DKK} a  new mechanism of PBH creation was proposed, the first 
ones, where inflation was invoked for PBH creation. This allowed to create PBHs  with masses
well above those predicted in the canonical approach, up to $10^5 M_\odot$ or even larger
than that.

A great achievement of these works was the prediction of the very simple
log-normal mass spectrum of PBHs:
\be
\frac{dN}{dM} = \mu^2 \exp{[-\gamma \ln^2 (M/M_0)], }
\label{dN-dM}
\ee
where $M_0$ should be equal to the mass in the cosmological horizon at the QCD phase transition.
A simple order of magnitude estimate gives
${ M_0 = M_{hor}  \approx 10 M_\odot}$, as estimated in our paper~\cite{AD-KP-M0} for negligibly 
small baryonic chemical potential,  i.e. for chemical potential of quarks.  
At larger chemical potential the 
temperature of the phase transition is  $T_{pt}$ is smaller and $M_{hor}$ is larger. 
As shown by the  lattice calculations~\cite{aref} for quark chemical potential $\mu/T \approx 0.3$,
 $M_0 = 17 M_\odot$. 
 This is exactly the value obtained by the best
fit to the data on gravitational waves registered by LIGO-Virgo-KAGRA, performed in 
the work~\cite{dkmppss}, where more references on the subject can be found.

On the other hand, the alternative analysis, performed in ref.~\cite{PKM-23} 
based on LIGO-Virgo-Kagra (LVK) collaboration (GWTC-3 catalogue),
indicates that the chirp mass distribution of the LVK BH binaries can
be fit with two distinct and almost equal populations: (1) astrophysical mergings
from BH+BH formed in the modern Universe from evolution of massive binaries
and (2) mergings of binary PBHs with initial log-normal mass distribution. It is 
found that for the PBH central mass $M_0 \approx 30M_\odot$, $\gamma \approx 10$, 
the observed LVK chirp masses are almost insensitive to the assumed double PBH
formation model. To comply with the observed LVK BH+BH merging rate, the
CDM PBH mass fraction should be $f_{pbh} \sim  10^{-3}$ but can be higher if PBH clustering
is taken into account. It is shown that almost equal
populations of astrophysical binary BHs from massive binary evolution and binary
PBHs with log-normal mass spectrum can describe both the observed chirp mass
distribution and effective spin – mass ratio anti-correlation of the LVK binary BHs.

Anyhow the fact that the proposed in refs.~\cite{DS,DKK} mechanism successfully 
described the observed features of PBHs
hints that the other prediction of these works, namely the prediction 
on galactic antimatter could also be valid.

\subsection{Mechanism of antimatter creation in the Galaxy \label{ss-mech}}

In papers~\cite{DS,DKK} a rather unusual way of PBH formation in comparison to
other approaches has been worked out. 
Firstly, it was assumed that the basis of PBH formation was created at the
cosmological inflationary stage. This was the first paper where such assumption has been made.
Now this is commonly used. Inflation allows to create PBH with very high masses, 
impossible with previously considered mechanisms. 

Another assumption is the highly essential role of a scalar field 
$\chi$ with nonzero baryonic number. 
Such field naturally appears in supersymmetric (SUSY)
baryogenesis scenario proposed by Affleck and Dine~\cite{aff-dine}.
In contrast to other mechanisms the cosmological baryon to photon ratio
generated by Affleck and Dine mechanism,
$\beta = N_B/N_\gamma$ is huge, of order unity, much larger than the
observed $\beta_{obs} = 6\times 10^{-10} $. Special efforts should be made
to suppress it down to the observed value, diminishing the initial value of $\chi$.

Another prediction of high energy SUSY models
is an existence of flat directions in the $\chi$-potential, either quartic (self-interaction):
\be
U_\lambda (\chi) = \lambda |\chi|^4 \left[1 -\cos (4\theta) \right], 
\label{U-lambda}
\ee
or quadratic, i.e. the mass term, $U_m = m^2 \chi^2 + m^{*2} \chi^{*2}$, and $m= |m| e^{i\alpha}$:
\be
U_m = m_\chi^2 |\chi|^2 \left[ 1 - \cos (2\theta +\alpha) \right] '
\label{U-m}
\ee
where $\chi = | \chi |^4\,\exp (i\theta) $. Here $\theta$ is the phase angle of $\chi$ field in the
complex plane $\left [\Re \chi, \Im \chi \right]$. 
If $\alpha \neq 0$, charge symmetry C/CP is broken.

As shown in several papers~\cite{VF,lin-dif,aas-dif,AD-DP}, at inflationary stage the average value 
of a real field $\chi$ 
rises with time as $\langle \chi^2 \rangle \sim 8 \rho\,t/(3H)$, where $\rho$ is the
cosmological energy density and $H$ is the Hubble  parameter, which is assumed to be 
larger than the mass of $\chi$. In our case of a complex field $\chi$
the average value of $\chi^2$ rises along the flat 
directions of the potential. For larger values of $\chi$ it rises along the flat directions 
of the self-interaction potential at $\cos (4\theta) =1$ and for smaller $\chi$ the rise goes along 
the mass term at $\cos (2\theta +\alpha)= 1$. Generally these flat directions
do not coincide for non-zero $\alpha$. 
In other words bosons $\chi$ condence via rising quantum 
fluctuation along flat directions of its potential. For large values of $\chi$ the 
quartic potential (\ref{U-lambda}) dominated
and $\chi^2$ could reach large values along of one of the directions with 
$\cos (4\theta) = 1 $.

Note that there is sign difference between the papers~\cite{VF,lin-dif,aas-dif}, 
and the later one~\cite{AD-DP}, though the absolute values are the same.
In paper~\cite{AD-DP}
the calculations of the average value of  $\langle\chi^2\rangle$ have been done by solution of the 
corresponding equation of motion, while in three previous ones by a regularization
of the infrared divergent quantity $\langle \chi^2 \rangle$. Solution of equation of motion is 
surely more reliable.

In GUT SUSY theories baryonic number is naturally non-conserved, because 
of generic non-invariance of  $U(\chi)$ with respect to the phase rotation, 
which is indeed demonstrated by the behavior of $U_\lambda$ (\ref{U-lambda}) 
or $U_m$ (\ref{U-m}).

When inflation is over or slows down to $H<m$, the field $\chi$
was far away from the origin $\chi =0$
and when inflation ends, it started
to evolve down to the equilibrium point, 
according to equation of motion that formally 
coincides with the equation of motion of a point-like particle in Newtonian mechanics:
\be
\ddot \chi +3H\dot \chi +U' (\chi) = 0.
\label{ddot-chi}
\ee
Baryonic charge of $ \chi$ is given by the expression:
\be
B_\chi =\dot\theta |\chi|^2 .
\label{B-chi}
\ee
It is analogous to mechanical angular momentum in complex $\chi$ plane. 

If $\lambda$ and  { $m$, are both non-zero,
the angular momentum, or what is the same, baryonic charge density, could be generated by a 
different direction of the quartic and quadratic valleys at low ${\chi}$.
CP would be broken, if the relative phase of ${\lambda}$ and 
$ m$ is non-zero, otherwise one can
``phase rotate'' $\bm \chi$ and arrive to real coefficients. 
{CP-odd phase ${\alpha}$ could be small but non-vanishing and both baryonic and 
antibaryonic domains might be  formed} {with possible dominance of one of them.}
The decay of ${{\chi}}$ would transfer its baryonic charge to that of quarks in B-conserving process.
Coupling to fermions may break CP, though not necessarily.
{Matter and antimatter objects may exist but globally is natural to expect that the net baryonic number 
density ${ B\neq 0}$.}
 
In refs.~\cite{DS,DKK} the following coupling of the  Affleck-Dine field $ \chi$  to inflaton $\Phi$
is proposed:
\be 
U = {g|\chi|^2 (\Phi -\Phi_1)^2}  +
\lambda |\chi|^4 \,\ln \left( \frac{|\chi|^2 }{\sigma^2 } \right)
+\lambda (\chi^4 + h.c. ) + 
(m^2 \chi^2 + h.c.). \,\,\,\,\,\,\,\,\,\,\,\,\,\,\,\,\,\,\,\,
\label{U-of-Phi}
\ee
The last two terms in this expression are the above mentioned quartic and quadratic
potential and the log-term is the Coleman-Weinberg effective potential~\cite{CW}, 
induced by one loop corrections to $\chi$ self-interactions. 

The first term describing the coupling of $\chi$ to inflation $\Phi$ is "invented" in our 
works~\cite{DS,DKK} for creation bubbles with a large baryonic number density
during relatively short time at 
inflationary stage. It is the general renormalizable coupling of two scalar fields.
The only fine tuning is that the value $\Phi_1$ is reached by the inflaton during inflation
when still considerable inflation proceeds, about 30-40 e-folding.

When $\Phi$ is close to $\Phi_1$, the window to the flat directions is open, 
the effective mass of $\chi$ along flat direction is zero and $\chi$
could reach quite large values,} according to quantum diffusion
equation derived by Starobinsky~\cite{Star-dif-eq}, generalized to a complex field ${\chi}$.
A large $\chi$ could generate cosmological baryon asymmetry of order unity.

If the window to flat direction, when $\Phi \approx \Phi_1$ 
is open only during a sifficienntly short
period, cosmologically small but possibly astronomically large bubbles with
high $\beta $ could be created, occupying a small fraction of the universe, while
the rest of the universe populated by small $\chi$, has normal $\beta = 6\cdot 10^{-10}$. This
mechanism of massive PBH formation is quite different from all others existing in the literature. The
fundament of PBH creation is build at inflation by making large
isocurvature fluctuations at relatively small scales, with practically vanishing density perturbations.
Initial isocurvature perturbations are hidden in the chemical content of massless
quarks. 

Density perturbations are generated much later, after the QCD phase transition, when massless quarks
turn into massive baryons and the bubbles with large baryonic number density became heavy and
could turn into primordial black holes. The emerging universe looks like a piece of Swiss cheese, 
"upside down", where holes are high baryonic density objects occupying a minor fraction of the 
universe space. The bubbles with smaller mass could form stellar-like objects, consisting either of
matter or antimatter more or less in equal share.


\section{Observations of galactic antimatter \label{s-anti-obs}}

Prediction of possible antimatter population in our Galaxy, was met by the community with the
"grain of salt" to say it mildly. But recent observations strongly indicate that there is indeed
some primordial antimatter in the Galaxy that definitely is not of secondary creation that could be
achieved through cosmic ray interactions with the interstellar matter.

\subsection{ Anti-evidence: cosmic positrons.} \label{ss-positrons}
 
Existence of rich populations of positrons in the Galaxy was noticed long ago through 
the observations of 511 keV gamma ray 
line (see~\cite{anti-e1, anti-e2, anti-e3} and references therein) with the flux 
\be\label{flux} {{
\Phi_{511 \; {\rm keV}} = { 1.07 \pm 0.03 \cdot 10^{-3} }\; 
{\rm photons \; cm^{-2} \, s^{-1}} .
}}
\ee
The width of the line is about 3 keV. 
The emission mostly goes from the  Galactic bulge and  at much lower level from the disk.
This unambiguously indicates the frequent annihilation of nonrelativistic $e^+ e^-$ pairs in the Galactic bulge with the rate~\cite{anti-e1}
\be \label{Neebulge}
\dot N_{ee}^\text{bulge} \sim 10^{43} \text{ s}^{-1}.
\ee

Note that one of the brightest X-ray  sources in the region around the Galactic Center   
got the name Great Annihilator~\cite{gr-ann}. Possibly it is a microquasar first detected in soft X-rays by the Einstein 
Observatory~\cite{Ein-observ} and later detected in hard X-rays by the space observatory ``Granat''~\cite{granat}.

There is no commonly accepted point of view on the origin of the cosmic positrons. The conventional hypothesis that positrons are created in strong 
magnetic fields of pulsars is at odds with the AMS data~\cite{AMS-19}. 
{However, this conclusion  is questioned in ref.~\cite{kachel} where it is shown that
{these features could be consistently explained by a nearby source which was active 
$\sim 2$ Myr ago and has injected  $(1-2)\times10^{50}$ erg in cosmic rays.

A competing option is that positrons are created by the Schwinger process at the horizon of small black holes with masses  $\gtrsim10^{20}$ g as is
suggested in ref.~\cite{BDP} and discussed in more detail in ref.~\cite{AD-AR}.

One more possibility that is closer to the spirit of this talk
is that positrons are primordial, produced in the early universe in relatively small antimatter 
domains~\cite{DS, DKK}. 

\subsection{Anti-evidence: cosmic antinuclei.} \label{ss-antinuc}

In 2018 AMS-02 announced possible observation of six
$\bm{\overline{He}^3}$ and two $\bm{\overline{He}^4}$~\cite{choutko,ting-CERN}.
Accumulated by 2022 data~\cite{anti-nuc-AMS-1, anti-nuc-AMS-2}
contains some more events:
7 $\overline D$ (at energies $\lesssim 15$ GeV) and 9 $\overline{He}^4$ at ($E\sim 50$ GeV).
These numbers correspond roughly speaking  to $\bm{\overline{He}/He \sim10^{-9}}$.
This number is much larger than the expected number of
$\overline{He}^4$, if it were created in cosmic ray collisions.
It is possible that the total flux of anti-helium is even much higher because low energy 
$\bm{\overline{He}}$ may escape registration in AMS.

The probability of the secondary creation of different antinuclei was estimated in ref.~\cite{cosm-anti-nuc}.
As argued in this work,  anti-deuterium could be  
most efficiently produced in the collisions ${\bar p\,p}$ or
${\bar p\, He}$ that can create the flux
${\sim 10^{-7} /m^{2}/ s^{-1}} $/steradian/GeV/neutron),
i.e. 5 orders of magnitude below the observed flux of antiprotons.
Antihelium could be created in the similar reactions and 
the fluxes of  ${\overline{He}^3}$ and ${\overline{He}^4}$, that
could be created in 
cosmic rays would respectively be 4 and 8
orders of magnitude smaller than the flux of the secondary created anti-D.

According to the works~\cite{DS,DKK}, antinuclei should be
primordial i.e. created in the very early universe during  
big bang nucleosynthesis (BBN) inside antimatter bubbles with high baryon density. 
However, the standard anti-BBN surely does not help, since normally BBN gives 75\% of hydrogen, 25\% of helium-4,
and a minor fraction of deuterium, at the level a few times $10^{-5}$,
in a huge contrast to the observed ratio of anti-deuterium to anti-helium which is of order unity. 
The same problem exists for the ratio
of $\overline{He}^3$ to $\overline{He}^4$, that is also of order unity instead of the standard  $\sim 3\times10^{-5}$.

If we assume that 
in the model of~\cite{DS,DKK}  the abundances of anti-D and anti-He are determined by normal BBN with large baryon-to-photon
ratio $\beta\sim1$, the problem would be even more pronounced, because amount of deuterium and helium-3 would be negligibly small,
even much less than $10^{-5}$.
On the other hand in our scenario  formation of primordial elements takes place inside non-expanding compact 
stellar-like objects with practically fixed temperature. If the temperature is sufficiently high, this so called BBN may stop 
with almost equal abundances of D and He. One can see that looking at 
abundances of light elements at a function of temperature. 
{If it is so, antistars may have equal amount of $\overline{D}$ and $\overline{He}$

\subsection{Anti-evidence: antistars in the Galaxy.} \label{ss-antistar}

A striking announcement of the 
possible discovery of anti-stars in the Galaxy was made in ref.~\cite{antistars}.
The catalog 14 antistar candidates was identified, not associated with any objects belonging 
to established gamma-ray source classes and with a spectrum compatible with baryon-antibaryon annihilation.

In ref.~\cite{bbbdp} a supplementary method of antistar identification was proposed.
Namely, in the process of the interaction of neutral atmospheres or winds from 
antistars with ionised interstellar gas, the hadronic annihilation 
will be preceded by the formation of excited $\bm{p \bar p}$
and $He {\bar p}$ atoms. These atoms rapidly cascade down to low levels prior to 
annihilation giving rise to a series of narrow lines which can be associated with the hadronic 
annihilation gamma-ray emission. The most significant are L (3p-2p) 1.73 keV line (yield more 
than 90\%) from ${p \bar p}$ atoms, and M (4-3) 4.86 keV (yield $\sim 60$\%) and L (3-2) 11.13 
keV (yield about 25\%) lines from $He^4 \bar p$ atoms. These lines can be probed in dedicated 
observations by forthcoming sensitive X-ray spectroscopic missions XRISM and Athena and in 
wide-field X-ray surveys like SRG/eROSITA all-sky survey.

The unexpectedly high flux of antinuclei~\cite{anti-nuc-AMS-1, anti-nuc-AMS-2} and 
an observation of antistars in the Galaxy~\cite{antistars} 
strongly supports the hypothesis, of ref.~\cite{bpbbd}, 
that antihelium cosmic rays are created by antistars.
It is shown that the flux of antihelium cosmic rays
reported by the AMS-02 experiment can be explained by 
{Galactic anti-nova outbursts, thermonuclear
anti-SN Ia explosions, a collection of flaring antistars, or an extragalactic source} with abundances not
violating existing gamma-ray and microlensing constraints on the antistar population.

As  argued in ref.~\cite{bpbbd}, a minor population of antistars in galaxies that has been predicted by some of non-standard
models of baryogenesis and nucleosynthesis in the early Universe, and their 
presence is not yet excluded
by the currently available observations. Detection of an unusually high abundance of antinuclei in
cosmic rays can probe the modified baryogenesis scenarios in the early Universe. 

 {Very powerful gamma-ray burster (GRB) was reported in ref.~\cite{star-annih}
 { This event got the nickname: the Brightest Of All Time or the { BOAT} }
  This extremely  strong GRB occurred in October 2022. 
{A bright megaelectronvolt emission line was observed,}
that appeared 280 seconds after the GRB began and then rapidly faded away while shifting to lower energies. 

{Usually the gamma-ray spectra of GRBs consist of a smooth continuum without absorption or 
emission lines.} {The authors interpret this MeV 
line as having been produced by the annihilation of electron-positron pairs 
within the  relativistic jet produced by the GRB}
{possibly emerging from star-antistar annihilation !?}
Star-antistar collision, was discussed in the talk~\cite{AD-Aosta}.  
It may be a quasi-periodic process of a star-antistar direct contact, with explosion pushing 
them apart, and possible, but not necessary  return to each other by gravitational attraction.

According to the works~\cite{DS,DKK}, antinuclei should be
primordial i.e. created in the very early universe during  
big bang nucleosynthesis (BBN) inside antimatter bubbles with high baryon density. 
However, the standard anti-BBN surely does not help, since normally BBN gives 75\% of hydrogen, 25\% of helium-4,
and a minor fraction of deuterium, at the level a few times $10^{-5}$,
in a huge contrast to the observed ratio of anti-deuterium to anti-helium which is of order unity. 
The same problem exists for the ratio
of $\overline{He}^3$ to $\overline{He}^4$, that is also of order unity instead of the standard  $\sim 3\times10^{-5}$.

If we assume that 
in the model of~\cite{DS,DKK}  the abundances of anti-D and anti-He are determined by normal BBN with large 
baryon-to-photon
ratio $\beta\sim1$, the problem would be even more pronounced, because amount of deuterium and helium-3 would be 
negligibly small,
even much less than $10^{-5}$.
On the other hand in our scenario  formation of primordial elements takes place inside non-expanding compact 
stellar-like objects with practically fixed temperature. If the temperature is sufficiently high, this so called BBN may stop 
with almost equal abundances of D and He. One can see that looking at 
abundances of light elements at a function of temperature. 
{If it is so, antistars may have equal amount of $\overline{D}$ and $\overline{He}$.


\section{Conclusion}

Outcome of the DS/DKK mechanism:
\begin{itemize}
\item
PBHs with log-normal mass spectrum - confirmed by the observations!
\item
{Compact stellar-like objects, similar to cores of red giants.}
\item
{Disperse hydrogen and helium clouds  with (much) higher than average $\bm{n_B}$ density.} 
\item
Strange stars with unusual chemistry and velocity.
\item
{${\beta}$ may be negative leading to creation of
(compact?) antistars which could survive annihilation despite being submerged into the 
homogeneous baryonic background.}
\item
Extremely old stars could exist  and indeed they are observed, even,
"older than universe star"  is found; its prehistoric age is mimicked by the unusual initial chemistry.  

\end{itemize}

{The mechanism of PBH creation pretty well verified by the data on the BH mass spectrum and 
on existence of antimatter in the Galaxy, especially of antistars. So we may expect that it indeed solves
the problems created by HST and JWST.}

Thus we may conclude that canonical  $\Lambda$CDM cosmology is saved by PBHs.}
{Antimatter in our backyard is predicted and found.}
 
 \section*{Acknowledgement}
 The work was supported by the state funding for neutrino physics FSUS-2025-0019.


\end{document}